\input{aipcheck}

\documentclass[final
    ,final            
  ]
  {aipproc}
\usepackage{amsmath}
\usepackage{amsfonts}
\usepackage{amssymb}
\usepackage{xcolor}
\usepackage{color}
\usepackage[english]{babel}
\usepackage{graphics}
\usepackage{graphicx}

\layoutstyle{8x11double}

\begin{document}

\title{Cosmology with a decaying vacuum}


\author{K. Urbanowski$^{1}$\footnote{e--mail: K.Urbanowski@proton.if.uz.zgora.pl}}{
    address={University of Zielona G\'{o}ra, Institute of Physics,
ul. Prof. Z. Szafrana 4a, 65--516 Zielona G\'{o}ra, Poland}
}

\author{M. Szyd{\l}owski$^{2}$\footnote{e--mail: marek.szydlowski@uj.edu.pl}}{
    address={  Jagiellonian University, Astronomical Observatory, ul. Orla 171, 30--244 Krak\'{o}w, Poland}
}

\begin{abstract}
Properties of  unstable false vacuum states are analyzed from the point of view of the quantum theory of unstable states. Some of false vacuum states survive up to times when their survival probability has a non-exponential form. At times much latter than the transition time, when contributions to the survival probability of its exponential and non-exponential parts are comparable, the survival probability as a function of time t has an inverse power-like form.  We show that at this time region the instantaneous energy of the false vacuum states tends to the energy of the true vacuum state as $1/t^{2}$ for $t \to \infty$.
\end{abstract}

\maketitle


\section{Introduction}

A discussion of the decay of the false vacuum was began in two pioneer
papers by Coleman and, Callan and Coleman \cite{Coleman,Callan}.
\textit{"The fate of the false vacuum"} was discussed there,
namely the unstability of a physical system in
a state which is not an absolute minimum of its energy
density, and which is separated from the minimum by
an effective potential barrier. It was shown, in those papers,
that even if the state of the early Universe is too
cold to activate a "{\it thermal}" transition (via thermal
fluctuations) to the lowest energy (i.e. {\it "true vacuum"}) state,
a quantum decay from the false vacuum to the true vacuum
may still be possible through a barrier penetration
via macroscopic quantum tunneling.
Not long ago, the decay of the false vacuum state in a cosmological
context has attracted renewed interest, especially
in view of its possible relevance in the process of
tunneling among the many vacuum states of the string
landscape (a set of vacua in the low energy
approximation of string theory).

Since the work of Khalfin  \cite{Khalfin}
it is known that for long times compared to the characteristic decay time
of an unstable state (when the decay law has an exponential form), the survival probability of such
states is no longer described by an exponential function of time $t$ but it decreases as  $t \rightarrow \infty $ more slowly
than any exponential function of $t$.
Krauss and Dent analyzing a false vacuum decay \cite{Krauss,Winitzki}
pointed out that in eternal inflation, even though regions of false vacua by assumption
should  decay exponentially, gravitational effects force space in a region that has not decayed yet
to grow exponentially fast. This effect causes that many false vacuum regions
can survive up to the times much later than times when the exponential decay law
holds. In the mentioned paper by Krauss and Dent the attention
was focused on the possible
behavior of the unstable false vacuum at very late times, where deviations from the exponential
decay law become to be dominat.
In our paper the attention will
be focussed on
properties of decaying false vacuum states from the point of view
of the quantum theory of unstable states evolving in time and decaying.
It will be shown that at late times
the instantaneous energy
of the false vacuum states tends to the
energy of the true vacuum state as
$1/t^{2}$ for $t \to \infty$.
This means that in the case of such false vacuum states the cosmological constant
becomes time depending.
Next effective cosmological model with obtained decaying Lambda term
will be studied in details.

\section{Unstable states in short}

If $|M\rangle$ is an initial unstable
state then the survival probability, ${\cal P}(t)$, equals
${\cal P}(t) = |A(t)|^{2}$,
where $A(t)$ is the survival amplitude,
$A(t) = \langle M|M;t\rangle,\;\;\;{\rm and}\;\; \; a(0) = 1$,
and
$[|M;t\rangle =
\exp\,[-itH]\,|M\rangle$,
$H$ is the total Hamiltonian of the system under considerations.
The spectrum, $\sigma(H)$, of $H$ is assumed to be bounded from below, $\sigma(H) =[E_{min},\infty)$
and $E_{min} > -\infty$.

From basic principles of quantum theory it is known that the
amplitude $A(t)$, and thus the decay law ${\cal P}_{M}(t)$ of the
unstable state $|M\rangle$, are completely determined by the
density of the energy distribution function $\omega({ E})$ for the system
in this state
\begin{equation}
A(t) = \int_{Spec.(H)} \omega({ E})\;
e^{\textstyle{-\frac{i}{\hbar}\,{ E}\,t}}\,d{ E}.
\label{a-spec}
\end{equation}
where
$\omega({E}) \geq 0$  and $\omega ({ E}) = 0$ for $E < E_{min}$.
From this last condition and from the Paley--Wiener
Theorem it follows that there must be (see \cite{Khalfin})
$|A(t)| \; \geq \; A_{1}\,\exp[- A_{2} \,t^{q}]$,
for $|t| \rightarrow \infty$. Here $A_{1} > 0,\,A_{2}> 0$ and $ 0 < q < 1$.

This means that the decay law ${\cal P}_{M}(t)$ of unstable
states decaying in the vacuum can not be described by
an exponential function of time $t$ if time $t$ is suitably long, $t
\rightarrow \infty$, and that for these lengths of time ${\cal
P}_{M}(t)$ tends to zero as $t \rightarrow \infty$  more slowly
than any exponential function of $t$.

The analysis of the models of
the decay processes shows that
${\cal P}_{M}(t) \simeq
\exp[- \frac{\Gamma_{M}t}{\hbar}]$,
 (where $\Gamma_{M}$ is the decay rate of the state $|M \rangle$),
to an very high accuracy  at the canonical decay times $t$:
From $t$ suitably later than the initial instant $t_{0}$
up to
$ t \gg \tau_{M} = \hbar / {\Gamma_{M}}$
and smaller than $t = T$, where $T$ is the crossover time and denotes the
time $t$ for which the non--exponential deviations of $A(t)$
begin to dominate.

In general,
in the case of quasi--stationary (metastable) states it is convenient to express $A(t)$ in the
following form
\begin{equation}
A(t) = A_{exp}(t) + A_{non}(t), \label{a-exp+}
\end{equation}
where $A_{exp}(t)$ is the exponential part of $A(t)$, that is
$A_{exp}(t) =
\exp[-it(E_{M} - \frac{i}{2}\,\Gamma_{M})] $,
($E_{M}$ is the energy of the system in the state $|M\rangle$ measured at the canonical decay times,
$N$ is the normalization constant), and $A_{non}(t)$ is the
non--exponential part of $A(t)$.
For times $t \sim \tau_{M}$:
$|A_{exp}(t)| \gg |A_{non}(t)|$,

The crossover time $T$
can be found by solving the following equation,
\begin{equation}
|A_{exp}(t)|^{\,2} = |A_{non}(t)|^{\,2}.
\end{equation}
The amplitude $A_{non}(t)$ exhibits inverse
power--law behavior at the late time region: $t \gg T$.
Indeed,  the integral representation (\ref{a-spec}) of $A(t)$ means that $A(t)$ is
the Fourier transform of the energy distribution function $\omega(E)$. Using this fact we can find
asymptotic form of $A(t)$ for $t \rightarrow \infty$. Results are rigorous.

Let us consider
$\omega ({E})$ having universal and general form. Namely
let $\omega ({ E})$ be of the form
\begin{equation}
\omega ({ E}) = ( { E} - { E}_{min})^{\lambda}\;\eta ({ E})\; \in \; L_{1}(-\infty, \infty),
\label{omega-eta}
\end{equation}
where $0 < \lambda < 1$ and it is assumed that $\eta (E_{min}) > 0$ and derivatives $\eta^{(k)}({ E})$,
($k= 0,1,\ldots, n$),  
exist and they are continuous
in $[{E}_{min}, \infty)$, and  limits 
$\lim_{{ E} \rightarrow {E}_{min}+}\;\eta^{(k)}({ E})$ exist,
$\lim_{{ E} \rightarrow \infty}\;( { E} - { E}_{min})^{\lambda}\,\eta^{(k)}({ E}) = 0$
for all above mentioned $k$, then
\begin{eqnarray}
A(t) & \begin{array}{c}
          {} \\
          \sim \\
          \scriptstyle{t \rightarrow \infty}
        \end{array} &
        (-1)\,e^{\textstyle{-\frac{i}{\hbar}{ E}_{min} t}}\;
        \Big[
        \Big(- \frac{i\hbar}{t}\Big)^{\lambda + 1}  \times \nonumber \\
         && \times \; \Gamma(\lambda + 1)\;\eta_{0}\; \label{a-eta} \\
        && +\;\lambda\,\Big(- \frac{i\hbar}{t}\Big)^{\lambda + 2} \times \nonumber \\
        && \times \; \Gamma(\lambda + 2)\;\eta_{0}^{(1)}\;+\;\ldots
        \Big]
         = A_{non}(t).
            \nonumber
\end{eqnarray}

\section{Energy and decay rate}

The amplitude $A(t)$ contains information about
the decay law ${\cal P}_{M}(t)$ of the state $|M\rangle$, that
is about the decay rate $\Gamma_{M}$ of this state, as well
as the energy ${E}_{M}$ of the system in this state.
This information can be extracted from $A(t)$. Indeed if
$|M\rangle$ is an unstable (a quasi--stationary) state then
$A(t)  \cong
\exp[\frac{i}{\hbar}({ E}_{M} -
\frac{i}{2} \Gamma_{M})\,t ] $.
So, there is
\begin{equation}
{E}_{M} - \frac{i}{2} \Gamma_{M} \equiv i
\hbar\,\frac{\partial A(t)}{\partial t} \; \frac{1}{A(t)},
\label{E-iG}
\end{equation}
in the case of quasi--stationary states.

Taking the above into account one can define the "effective
Hamiltonian", $h_{M}$, for the one--dimensional subspace of
states ${\cal H}_{||}$ spanned by the normalized vector
$|M\rangle$ as follows
\begin{equation}
h_{M} \stackrel{\rm def}{=}  i \hbar\, \frac{\partial
A(t)}{\partial t} \; \frac{1}{A(t)}. \label{h}
\end{equation}
In general, $h_{M}$ can depend on time $t$, $h_{M}\equiv
h_{M}(t)$. One meets this effective Hamiltonian when one starts
with the Schr\"{o}dinger Equation
for the total state
space ${\cal H}$ and looks for the rigorous evolution equation for
the distinguished subspace of states ${\cal H}_{||} \subset {\cal
H}$. Details can be found in \cite{EPJD-2009} and in \cite{CEJP-2009}.
Thus,
one finds the following expressions for the
energy and the decay rate of the system in the state $|M\rangle$
under considerations, to be more precise for
the instantaneous energy ${\cal E}_{M}(t)$ and the  instantaneous decay rate,
$\gamma_{M}(t)$,
${\cal E}_{M}\equiv {\cal E}_{M}(t) = \Re\,(h_{M}(t))$,
$\gamma_{M} \equiv \gamma_{M}(t) = -\,2\,\Im\,(h_{M}(t))$,
where $\Re\,(z)$ and $\Im\,(z)$ denote the real and imaginary parts
of $z$ respectively.

Using (\ref{h})  one can find that
${\cal E}_{M} (t \sim \tau_{M}) \simeq { E}_{M} \neq {\cal E}_{M} (0)$,
$\gamma_{M}(t \sim \tau_{M}) \simeq  \Gamma_{M}$,
i. e., there is ${\cal E}_{M}(t)= E_{M}$ at the canonical decay time.

Starting from the  asymptotic expression
(\ref{a-eta}) for $A(t)$ and using (\ref{h})
after some algebra one finds for  times $t \gg T$  that
\begin{equation}
{h_{M}(t)\vline}_{\,t \rightarrow \infty} \simeq { E}_{min} + (-\,\frac{i\hbar}{t})\,c_{1} \,
+\,(-\,\frac{i\hbar}{t})^{2}\,c_{2} \,+\,\ldots, \label{h-infty-gen}
\end{equation}
where $ c_{i} = c_{i}^{\ast},\;\;i = 1,2,\ldots$; (coefficients $c_{i}$ depend on  $\omega (E)$).
This  last relation means that
\begin{eqnarray}
{\cal E}_{M}(t) &\simeq & E_{min} \, +  \,\frac{c_{2}}{t^{2}} \ldots, \;\;\;({\rm for}
\;\;t \gg T). \label{E(t)}\\
\end{eqnarray}
(Here $c_{2} > 0$). These properties take place for  all unstable states which survived up to times $t \gg T$.
From (\ref{E(t)}) it follows that
$\lim_{t \rightarrow \infty}\, {\cal E}_{M}(t) = E_{min}$.

\section{Cosmological applications}

Krauss and Dent in their paper \cite{Krauss}  mentioned earlier
made a hypothesis
that some false vacuum regions do survive well up to the time $T$ or  later.
Let $|M\rangle = | 0\rangle^{false}$,
be a false,$|0\rangle^{true}$ -- a true, vacuum states and  $E^{false}_{0}$ be the energy of a state corresponding to the false vacuum measured at the canonical decay time
and $E^{true}_{0}$ be the energy of true vacuum (i.e. the true ground state of the system).
As it is seen from the results presented in previous Section, the problem is that the energy of those false
vacuum regions which survived up to $T$ and much later differs from $E^{false}_{0}$, (see \cite{PRL-2011} and
references one can find therein).

Going from quantum mechanics to quantum field theory one should take into account  among others a volume factors so that survival probabilities per unit volume per unit time should be considered. The standard false vacuum decay calculations shows that the same volume factors should appear in both early and late time decay rate estimates (see \cite{Krauss} ). This means that the calculations of cross--over time $T$ can be applied to survival probabilities per unit
volume.  For the same reasons  within the quantum field theory the quantity ${\cal E}_{M}(t)$ can be replaced by  the energy per unit volume $\rho_{M}$ because these volume factors appear in the numerator and denominator of the formula (10) for $h_{M}(t)$.
Now, if  one assumes that $E^{true}_{0} \equiv E_{min}$ then one has for the energy  of the false vacuum state that at
$t \gg T$,
\begin{equation}
{\cal E}^{false}_{0}(t) \simeq E^{true}_{0}  + \frac{c_{2}}{t^{2}}\ldots \;\; \neq\,E^{false}_{0}.
\label{E-false-infty}
\end{equation}
This property of  the false vacuum states
means that
\begin{equation}
{\cal E}^{false}_{0}(t) \rightarrow E^{true}_{0} \;\;\; {\rm as}\;\;t \to \infty.
\label{E-false-lim}
\end{equation}

The basic physical factor forcing the wave function  $|M;t\rangle$ and thus the  amplitude $A(t)$
to exhibit inverse power law behavior at $t \gg T$ is a boundedness from below of  $\sigma (H)$. This means
that if this condition takes place and
$\int _{-\infty}^{+\infty}\omega(E)\,dE\,< \,\infty $,
then all  properties
of $A(t)$, including a form of the time--dependence at  $t \gg T$, are the  mathematical consequence of them both.
The same applies by (\ref{h}) to properties of $h_{M}(t)$ and concerns
the asymptotic form of $h_{M}(t)$ and
thus of ${\cal E}_{M}(t)$  and $\gamma_{M}(t)$ at $t \gg T$.

 Note that properties of $A(t)$ and $h_{M}(t)$ discussed above
do not take place when  $\sigma(H) = (-\infty, + \infty)$.
The late time behavior of the energy of the system in the false vacuum state,
\begin{equation}
{\cal E}^{false}_{0}(t) \simeq E^{true}_{0}  + \frac{c_{2}}{t^{2}}\ldots ,\;\;\; {\rm for}\;\;\; t \gg T,
\label{E-false-infty-1}
\end{equation}
or, of the energy density $\rho^{false}_{0}(t)$:
\begin{equation}
{\rho}^{false}_{0}(t) \simeq \rho^{true}_{0}  + \frac{d_{2}}{t^{2}}\ldots ,\;\;\; {\rm for}\;\;\; t \gg T,
\label{rho-false-infty-2}
\end{equation}
(where $d_{i}=d_{i}^{\ast}, \, i =,1,2,\ldots$),  is the pure quantum effect following from the basic principles of the quantum theory.
The standard relation is $\rho_{0}^{true} = \frac{\Lambda}{8 \pi G} =\rho_{bare}$.

\section{False vacuum cosmology}

Let us take into account properties of the false vacuum described by the last relations and
let us consider the cosmological model which is homogeneous and isotropic (possesses the Robertson - Walker metric) in which energy density of dark sector is given by the relation which consists of the bare cosmological constant $\Lambda$ and the term depending on the cosmological time $t$
\begin{equation}
\rho=\rho_{\text{de}}=\Lambda+\frac{\alpha^2}{t^2}. \label{rho-de}
\end{equation}
As it is well know the Friedmann-Robertson-Walker model has a first integral which does not depend on the pressure $p$ and is given in the form called the Friedmann equation
\begin{equation}
3H^2=\rho_{\text{de}}+\rho_{\text{m}}, \label{F-eq}
\end{equation}
where $H=\frac{d\ln a}{dt}$ is the Hubble parameter, 
$t$ is the cosmological time and $a=a(t)$ is a scale factor.

For simplicity we assume that space is flat and matter is in the form of dust ($p=0$), so
$\rho \equiv \rho_{m} =\rho_{\text{m},0}\; a^{-3} $
describes a dependence of the energy density of the matter on a scale factor or a redshift $z \colon 1+z=a^{-1}$.
Therefore  (\ref{rho-de}) and (\ref{F-eq})
give
\begin{equation}
\frac{\rho_{\text{m},0}a^{-3}}{3H^2}+\frac{\Lambda}{3H^2}+\frac{\alpha^2/t^2}{3H^2}\equiv 1.
\end{equation}
Next we introduce the state variables ($x$, $y$, $z$) defined as
$\sqrt{\Omega_{\text{m}}}= x \equiv \sqrt{\frac{\rho_{\text{m},0}a^{-3}}{3H^2}}$,
$\sqrt{\Omega_{\Lambda}}=y \equiv \sqrt{\frac{\Lambda}{3H^2}}$ and
$\sqrt{\Omega_\alpha}=z \equiv \sqrt{\frac{\alpha^2/t^2}{3H^2}}$.
The quantities {$x$}, {$y$}, {$z$} has the cosmological sense of {density parameters} for a {matter}, the {cosmological constant} and for the which mimics time dependence of the false decaying vacuum. The cosmological constant $\Lambda$ is assumed to be positive.
 Above variables, of course, satisfy the constrain $x^2+y^2+z^2=1$.
Using
this constrain
after some algebra  we obtain
for state variables $(x,y,z)$
dynamical system in the phase space $(x, y, z)\colon x^2+y^2+z^2=1$, (i.e. on the unit sphere), having the
following form:
\begin{align}
\label{eq:18}
\frac{dx}{d\tau} & = x\left(\frac{3}{2}x^2+\frac{\sqrt{3}}{\alpha}z^3-\frac{3}{2}\right) \\
\label{eq:19}
\frac{dy}{d\tau} & = y\left(\frac{3}{2}x^2+\frac{\sqrt{3}}{\alpha}z^3\right) \\
\label{eq:20}
\frac{dz}{d \tau} & = z\left(-\frac{\sqrt{3}}{2}z+\frac{3}{2}x^2+\frac{\sqrt{3}}{\alpha}z^3\right),
\end{align}
where $z=\pm\sqrt{1-x^2-y^2}$ and
the cosmological time $t$ is reparametrized to the new time parameter $\tau$ which is a monotonic function of $t$, i.e.,
$\tau \equiv \ln a $.
The above system has three 2D invariant submanifolds: $\{x=0\}$, $\{y=0\}$, $\{z=0\}$ and 
the following critical points:
\begin{enumerate}
	\item {$x=1$, $y=0$, $z=0$ (Einstein-de Sitter universe),}
	\item {$x=0$, $y=1$, $z=0$ (de Sitter stationary universe),}
	\item {$z=1$, $x=0$, $y=0$ (Milne universe),}
	\item {$y=0$, $x=\frac{1}{2}\sqrt{4-3\alpha^2}$, $z=\frac{1}{2} \sqrt{3\alpha^2}$ (Einstein-de Sitter dominated by the time-dependent part of the cosmological constant).}
\end{enumerate}

    If we eliminate one state variables from the constraint condition then model dynamics constitute the 2D dynamical system and then right-hand sides of the system (\ref{eq:18}) --- (\ref{eq:20}) are not given in the polynomial forms.
	In neighborhood of initial singularity the generic solution is representing $\Lambda$CDM universe dominating.
    There is a new critical point representing by a node and saddle point representing the Milne universe (without the horizon). In this case $a=a_0 t$ and we have an unstable invariant submanifold, which is 2D if $\alpha \gg 1$, and an unstable node if $\alpha \ll 1$. For the late time ($t\rightarrow\infty$) the trajectories of the system are going to the Sitter universe like for the FRW model with the bare cosmological constant.

From dynamical analysis of the system one can observe how changes the critical points as well as its type. There are present bifurcation as the value of $\alpha$ parameter changes. For example the bifurcation value of $\alpha_{\text{bif}}$ for the critical point (4) is: $\alpha_{\text{bif}}=\frac{2}{\sqrt{3}}$.

\begin{figure} [ht!]
\centering
\includegraphics[scale=0.5] {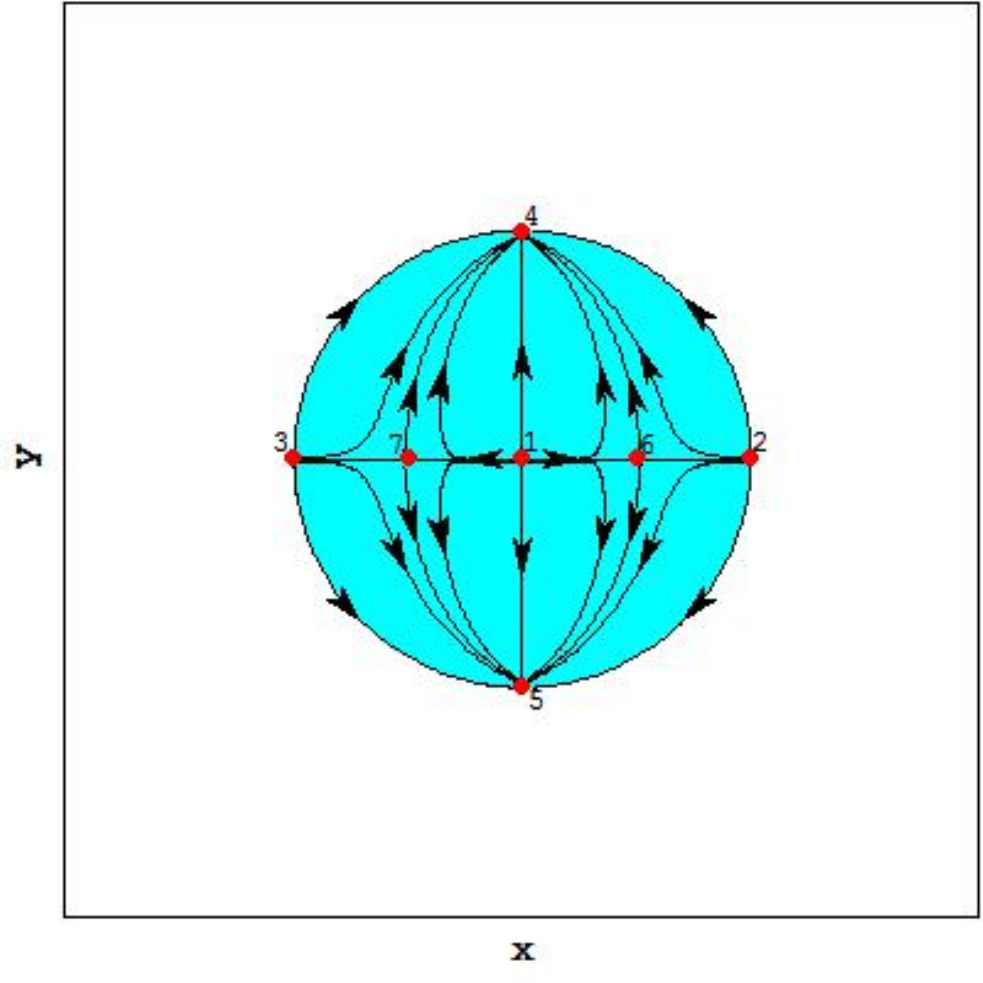}
\caption { A phase portrait of the system (\ref{eq:18})-(\ref{eq:20}) for $\alpha=1$. There are three critical points at a finite domain labeled as (7), (1), (6) and four critical points at the boundary $x^2+y^2=1$.
Notations: (1)  the Milne universe dominated by the time dependent part of the cosmological constant; (2)  the Einstein -- de Sitter universe dominated by matter; (4)  the de Sitter universe dominated by the bare $\Lambda$; (6)  the Einstein-de Sitter universe dominated by the time dependent part of the cosmological constant.}
\label{fig:1}
\end {figure}

\begin {figure} [ht!]
\centering
\includegraphics[scale=0.5] {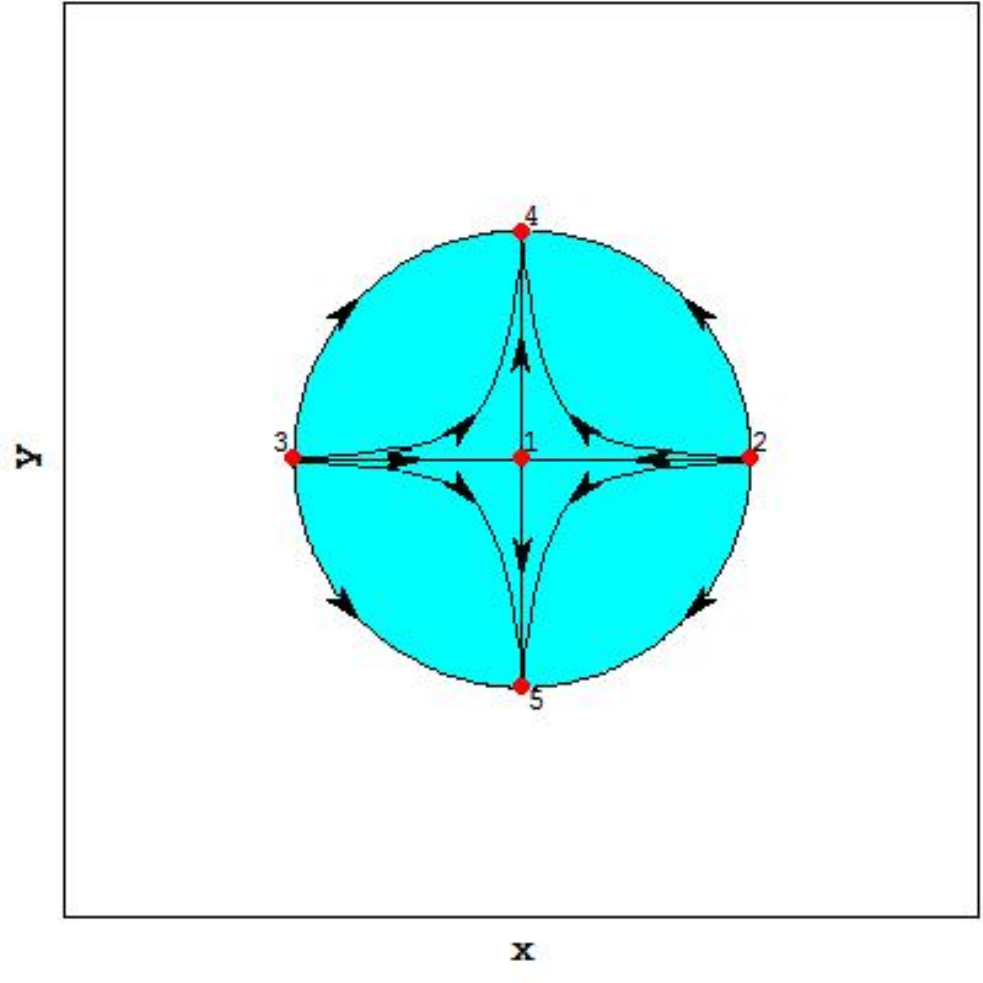}
\caption{A phase portrait of the system (\ref{eq:18})-(\ref{eq:20}) for $\alpha=2$.
 On the phase portrait there is one critical point at finite domain of saddle type and four critical points located at the boundary $\Omega_m+\Omega_{\Lambda}=1$.
}
\label{fig:2}
\end {figure}

\section*{Conclusions}

We have shown that at late times,  {$t \gg T$}, the energy of the false vacuum  behaves as follows:
\begin{equation}
{\cal E}^{false}_{0}(t) \simeq E^{true}_{0}  + \frac{c_{2}}{t^{2}}\ldots ,\;\;\; {\rm for}\;\;\; t \gg T,
\label{E-false-infty-aa}
\end{equation}
($c_{2} > 0$), and similarly the energy density {$\rho^{false}_{0}(t)$}:
\begin{equation}
{\rho}^{false}_{0}(t) \simeq \rho^{true}_{0}  + \frac{d_{2}}{t^{2}}\ldots ,\;\;\; {\rm for}\;\;\; t \gg T.
\label{rho-false-infty-bb}
\end{equation}

Using the above property of the false vacuum we have
 studied the dynamics of the flat cosmological FRW model with decaying false vacuum parameterized by the cosmological time.





\bibliographystyle{aipproc}   


\IfFileExists{\jobname.bbl}{}
 {\typeout{}
  \typeout{******************************************}
  \typeout{** Please run "bibtex \jobname" to optain}
  \typeout{** the bibliography and then re-run LaTeX}
  \typeout{** twice to fix the references!}
  \typeout{******************************************}
  \typeout{}
 }

\end{document}